\documentclass[prl,twocolumn,amsmath,amssymb,showpacs,superscriptaddress,floatfix]{revtex4}
\def\vc{\mathbf}
\newcommand{\pipp}[0]{\textsc{pip}$_2$}
\newcommand{\pipfrac}[0]{\phi_{\textsc{PIP}}}
\newcommand{\Rci}[0]{R_\mathrm{CI}}
\newcommand{\qpip}[0]{q_{\textsc{PIP}}}
\newcommand{\kT}[0]{k_\mathrm{B}T}

\newcommand{\lB}[0]{l_\mathrm{B}}
\newcommand{\capp}[0]{Ca$^{2+}$}
\usepackage{graphicx}
\begin{document}
\title{Electrostatic cluster formation in lipid monolayers}

\author{Wouter G. Ellenbroek}
\affiliation{Department of Physics and Astronomy, University of Pennsylvania, Philadelphia, PA 19104, USA}

\author{Yu-Hsiu Wang}
\affiliation{Department of Chemistry, University of Pennsylvania, Philadelphia, PA 19104, USA}

\author{David A. Christian}
\affiliation{Chemical \& Biomolecular Engineering, University of Pennsylvania, Philadelphia, PA 19104, USA}

\author{Dennis E. Discher}
\affiliation{Chemical \& Biomolecular Engineering, University of Pennsylvania, Philadelphia, PA 19104, USA}

\author{Paul A. Janmey}
\affiliation{Institute for Medicine and Engineering, University of Pennsylvania, Philadelphia, PA 19104, USA}
\affiliation{Departments of Physiology and Bioengineering, University of Pennsylvania, Philadelphia, PA 19104, USA}
\affiliation{Department of Physics and Astronomy, University of Pennsylvania, Philadelphia, PA 19104, USA}

\author{Andrea J. Liu}
\affiliation{Department of Physics and Astronomy, University of Pennsylvania, Philadelphia, PA 19104, USA}
\affiliation{Department of Chemistry, University of Pennsylvania, Philadelphia, PA 19104, USA}

\date{\today}
\begin{abstract}
We study phase separation in mixed monolayers of neutral and highly negatively charged
lipids, induced by the addition of divalent positively charged counterions.  We
find good agreement between experiments on mixtures of \textsc{pip}$_2$ and
\textsc{sopc} and simulations of a simplified model in which only the essential
electrostatic interactions are retained. Thus, our results support an
interpretation of \textsc{pip}$_2$ clustering as governed primarily by
electrostatic interactions, in which divalent ions such as calcium mediate an
effective attraction between like-charged lipids.  Surprisingly, the mediated
attractions are strong enough to give nearly complete phase separation, so that
clusters can even form when the overall concentration of \textsc{pip}$_2$ is low, as is
the case in the cell membrane.
\end{abstract}
\maketitle

One of the truisms of biology is that electrostatics play a relatively
unimportant role in determining structure on scales above the Debye screening
length, which is on the nanometer scale. Highly charged biomolecules such as
DNA~\cite{guldbrand86} and actin~\cite{tang96jbc} form an exception to this
rule by aggregating into large bundles in the presence of multivalent ions.
Another biomolecule that carries a very high negative charge density is the
membrane lipid \pipp.  Despite the fact that its concentration in the membrane
is extremely low (of order 1\%), this lipid plays a critical role in many
processes involving the cell membrane, including cell division~\cite{saul04},
exchange of chemicals with the environment through endocytosis and
exocytosis~\cite{martin01}, and cell motility~\cite{janmey94}.  Evidence exists
that \pipp\ forms clusters~\cite{levental09} at the sub-micron scale under
roughly physiological conditions.
It has been conjectured that this clustering is crucial to its effectiveness at
such low overall concentration~\cite{martinbelmonte07,richer09}. 

Various mechanisms for the clustering have been proposed, including
\pipp-protein interactions~\cite{mclaughlin05,laux00}, exclusion from
cholesterol-enriched ordered domains~\cite{levental09bcj,levental09}, and
hydrogen bonds~\cite{levental08jacs,redfern05}. However, recent
experiments showed that \pipp-clusters are induced simply by adding calcium or
other divalent ions~\cite{carvalho08,levental09}.
This raises the question of whether a purely electrostatic, counterion-mediated
mechanism is consistent with experimental observations of calcium-induced
\pipp-clustering.

In this letter, we study ion-induced clustering of negatively charged lipids as
a function of lipid charge and ion properties, both numerically and
experimentally.  We conduct simulations on a model designed to retain only the
most critical features of the electrostatics  and compare the results to
experiments on Langmuir monolayers of a mixture of \pipp\ with neutral lipids
with added divalent salts. We find semiquantitative agreement between the
simulations and experiments, with similar trends for the dependence on lipid
charge and ion size, suggesting that multivalent-ion-mediated attractions are
indeed responsible for the observed clustering.   

Counterion-mediated attractions are the collective result of a
near-cancellation of repulsive and attractive interactions between like and
unlike charges, respectively, in strongly correlated electrostatic systems.  They are not captured in mean-field theory~\cite{levin02} and are typically quite
weak~\cite{leeliu04}. Here, we show that counterion-mediated attractions are
surprisingly strong so that phase separation is nearly complete even at
reasonable values of the \pipp\ charge, implying that clustering can occur even
at very low \pipp\ concentration.

\begin{figure*}[!t]
\includegraphics[width=150mm]{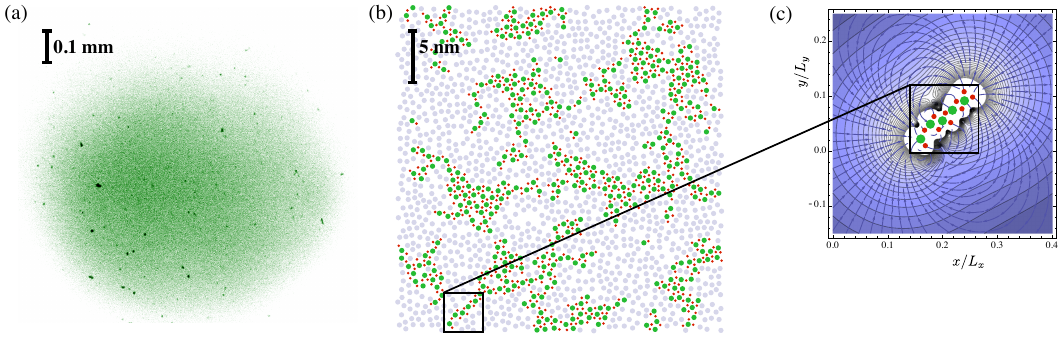}
\caption{Snapshots of the experiment (a) and simulation (b) at $\pipfrac=25\%$. (a)~The (inverted)
epifluorescence micrograph, taken 50 minutes after mixing 1\,mM CaCl$_2$
into the subphase (pH 7.4). \pipp-rich domains are shown as very dark green spots.
(b)~The simulation (\pipp-charge $\qpip=-4$, divalent ion radius $R_\mathrm{CI}=2\,$\AA) after 3.5\,ps of
coarsening. Charged and neutral lipids are drawn green and light grey, respectively,
and divalent ions that are close to the lipid monolayer are plotted in dark red.
(c)~Strength (shaded contours) and direction (streamlines) of the electric field
around a string-like domain (geometry taken from the simulation),
illustrating that further growth of the domain is
likely to occur at the end.}
\label{fig:snaps}
\end{figure*}
\emph{Experiments}---We look for phase separation using visual analysis of
fluorescence micrographs of mixed lipid monolayers prepared in a Langmuir
trough (Kibron.\ Inc.) and imaged on an inverted epifluorescence
microscope.  We use a molar fraction $\pipfrac$ of
L-$\alpha$-phosphatidylinositol-4,5-bisphosphate (\pipp, Avanti) in a monolayer
otherwise consisting of 1-stearoyl-2-oleoyl phosphatidylcholine
(\textsc{sopc}, Avanti). \textsc{sopc} is
zwitterionic, and known to be net-neutral over a wide range of pH.
Part of the \pipp\ (0.5 mol\% of the total lipid content) is replaced by a
fluorescently labeled analogue (Bodipy FL-\pipp, Echelon, Inc.). A lipid monolayer is formed
on a buffered subphase (10\,mM \textsc{hepes}, 100$\,\mu$M \textsc{edta}, 5\,mM
\textsc{dtt}) by addition of the lipids dissolved in a  2:1 chloroform/methanol mixture
to the air-water interface.
After formation, the monolayer is visualized by epifluorescence to
determine a baseline frequency of structural inhomogeneity. Divalent salts
CaCl$_2$ or MgCl$_2$ were added at 1\,mM to the subphase, followed by gentle
mixing to avoid disrupting the monolayer. We image the monolayers about
50 minutes after cation addition to allow sufficient time for domain
coarsening. The existence of bright spots at a higher frequency than baseline
serves to determine the existence of \pipp-rich clusters.

We perform this procedure for a range of $\pipfrac$-values and several pH
values: $3,\,4.5,\,6,\,7.4,\,9$. At these values of the pH, $\qpip$ is roughly
$-1.5,\,-2.7,\,-3.2,\,-4.2,\,-5.0$, respectively, based on acid dissociation
constants from Ref.~\cite{levental08bpj}.  However, because the ionization
state of \pipp\ may be influenced by various geometric and chemical
factors~\cite{levental08bpj}, we do not assume that these $\qpip$ values are
exact for our system.

\emph{Simulations}---We retain only the competition between electrostatic interactions and excluded volume repulsions by adopting a model in which both lipids and small ions are
represented as charged spheres
(radius $R_i$) with an excluded volume interaction given by the purely repulsive (truncated at its minimum and shifted)
Lennard-Jones potential (the WCA potential~\cite{wca71}).  Parametrized by an energy scale $\epsilon=\kT\equiv 1$ (our unit of energy) and length
scale $\sigma_{ij}=R_i+R_j$, this potential takes the following form as a function of
center-to-center distance $r_{ij}$,
$$
V_{\mathrm{WCA},ij}(r_{ij})=4\epsilon\left[\left(\frac{\sigma_{ij}}{r_{ij}}\right)^{12}-\left(\frac{\sigma_{ij}}{r_{ij}}\right)^6+\frac14\right],
$$
for $r_{ij}<2^{1/6}\sigma_{ij}$, and $V(r_{ij})=0$ otherwise. Note that
$\sigma_{ij}$ is the distance at which the potential equals $\kT$. $N=1600$ lipid
particles are confined to the $z=0$ plane, to mimic the effect of the
hydrophobic interaction that keeps them at the air-water interface. We use
$R_i=R_\mathrm{L}=3\,$\AA\ for the lipids and $R_i=\Rci=2\,$\AA\ for the small cations that are allowed to explore the entire simulation box. In a study of the dependence of the
clustering on cation size, we vary it in the range $0.5\,$\AA$\ \leq
\Rci\leq2.5\,$\AA. The box is periodic in $x$- and $y$-directions (size
$L_x=320\,$\AA$ \times L_y=320\,$\AA) and has hard walls at $z=0$ and $z=L_z=200\,$\AA. The typical distance between lipids in the monolayer at $z=0$ is therefore $8\,$\AA.

In addition, the charged spheres interact via the Coulomb interaction,
$V_{\mathrm{C},ij}=q_iq_j\lB/r_{ij}$, where we measure
charges $q$ in units of the proton charge. In room temperature water the
Bjerrum length $\lB\approx 7\,$\AA. 

We run molecular dynamics simulations using LAMMPS~\cite{lammps}, with a
Nos\'e-Hoover thermostat~\cite{hoover85} and PPPM for the long range Coulomb
interactions~\cite{hockney88}.

The strong Coulomb attraction between the anionic lipids and the small cations
allows them to bind at a distance of rougly $\sigma_{ij}$. The essence of
ion-mediated attractions is that these bonds are strong and long-lived enough
so that one or two counterions can draw together two lipids and be bound to
both simultaneously~\cite{cmamech}. Due to its coarse-grained nature,
our model underestimates the binding energy of such bonds in two ways. Firstly,
in real \pipp\ the negative charges are localized mainly in phosphate groups
that lie close to the surface of the molecule, so that the distance between the
phosphate groups and the cations is much closer than the typical $5\,$\AA\
allowed by our ``spherical'' lipids. To determine the effect of the
inter-charge distance on the binding energy, we performed two test calculations
with two ``lipids'' ($\qpip=-4$) and four divalent ions. We found that the
binding energy is 2.2 times larger for a structured lipid with two charges
($q=-2$) about $4\,$\AA\ apart, than for the spherical lipids in our model.
Secondly, \capp\ can be expected to lose some of its hydration shell when it
binds to a phosphate group, so that the effect of water as a dielectric medium
is partially eliminated. This further increases the net binding energies by an
unknown amount. As an estimate, we assume the binding energies produced by our
model interaction are a factor of 3 lower than those in the real system and
compensate for this by using a dielectric constant lower by a factor of 3;
i.e., we use a Bjerrum length of 21\,\AA.

\begin{figure}[t]
\includegraphics[width=86mm]{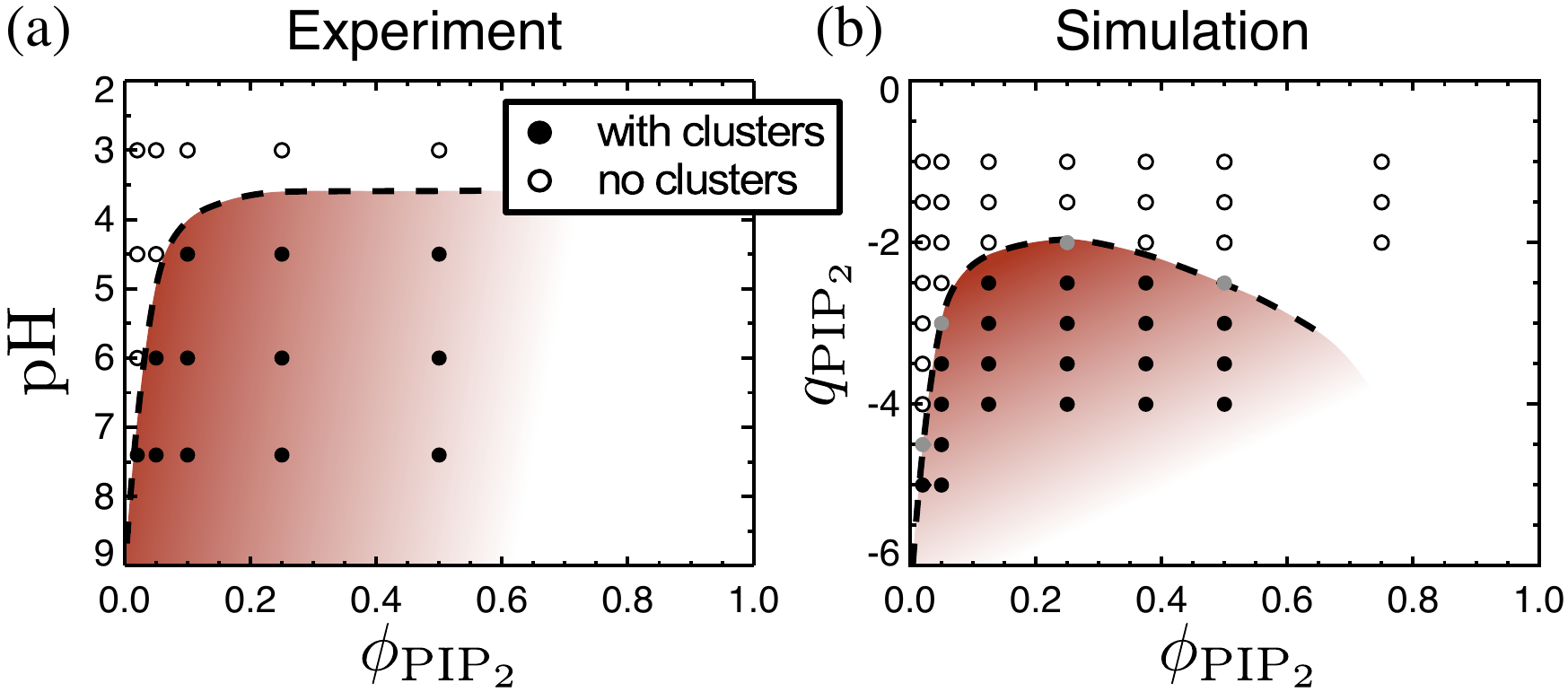}
\caption{Phase diagram (charge vs. \pipp-fraction) of a charged-neutral mixed
monolayer (a) in experiments, where the lipid charge is set by the pH, and (b)
in simulations (divalent ion radius $\Rci=2\,$\AA). The closed discs in the
shaded coexistence region indicate where clustering was
observed.  Open circles mark mixed samples, and grey discs are too close to the
boundary to determine their behavior with certainty.}
\label{fig:pd}
\end{figure}
\emph{Results}---At high \pipp-charge, for example at pH 7.4 in the experiment,
where $\qpip\approx-4.2$, or at
$\qpip=-4$ in the simulation, cluster formation is readily observed
(Fig.~\ref{fig:snaps}). The experimental image (Fig.~\ref{fig:snaps}a) shows
the larger-scale picture of bright fluorescent spots marking the regions of
large \pipp-concentration, while the simulation image (Fig.~\ref{fig:snaps}b)
shows still growing clusters at a length scale that is 1000
times smaller, after $3.5\,$ps of simulation time. As expected, the positions
of the condensed calcium ions (red discs in Fig.~\ref{fig:snaps}b) clearly
indicate their role in binding the charged lipids (green discs) together.

The morphology observed in the early stages of
coarsening in the simulations illustrates some particular features of
ion-mediated attractions that set them apart from simple attractive potentials. As
can be seen in Fig.~\ref{fig:snaps}b, the \pipp-rich clusters are often
irregularly shaped, and even string-like. This occurs because the
attraction is the net result of strong attractions (\pipp-\capp) and strong
repulsions (\pipp-\pipp\ and \capp-\capp) that can both be several tens of
$\kT$. Hence, a rearrangement of the lipids in a cluster typically involves
energy barriers that are much higher than the net attraction energies, so that
evolution towards more compact shapes is severely hindered kinetically. In the
earliest stages of coarsening, most domains are string-like, because for very
small clusters such linear arrangements have the lowest Coulomb energy. As the
domains grow, compact shapes become energetically favorable but are difficult
to reach for two reasons. First, once there is a string-like cluster, the
electric field in its neighborhood is focused towards the end of the string
(see Fig.~\ref{fig:snaps}c), which makes it more likely for the next lipid to
bind at the end, thus extending the string.  Second, the energy barrier for the
string to fold onto itself is quite high. Therefore, string-like domains can persist
even in the later stages of coarsening, as seen for example in the bottom left
of Fig.~\ref{fig:snaps}b. Such irregular domains have been seen
experimentally~\cite{levental09,mclaughlin05}.

To determine the conditions under which cluster formation occurs, a grid of
parameter values ($\pipfrac,\qpip$) was explored. Experimentally, we identify which
images have more bright spots than the baseline level (which we found to be
at most two spots per frame).
The resulting phase diagram is shown in Fig.~\ref{fig:pd}a, where
the region of cluster formation is shaded. In the simulations, we follow the
coarsening dynamics by keeping track of the static structure factor of the
charged lipids,

$$
S(\vc{k})=\frac1N\sum_{i,j}^N\exp[i\vc{k}\cdot(\vc{r}_i-\vc{r}_j)]~,
$$
where $N$ is the number of \pipp-particles.  As a function of
$k\equiv|\vc{k}|$, a maximum in this function at $k=k_\mathrm{peak}$ indicates
that the \pipp-postions are developing structure at a length scale
$2\pi/k_\mathrm{peak}$.  For the more pronounced cases of cluster formation
(deep in the phase-separated regime), we followed this peak as a function of
time and verified that it scales with time as $k_\mathrm{peak}\sim t^{-1/3}$,
consistent with the general theory of coarsening of a binary fluid
mixture~\cite{lifshitz61}.   Thus, even though the counterion-mediated origin
of phase separation yields irregularly shaped clusters instead of circular
ones, this does not seem to affect the kinetics of coarsening. In the
phase diagram in Fig.~\ref{fig:pd}b, all parameter values ($\pipfrac,\qpip$)
for which an appreciable peak appears that approaches $k_\mathrm{peak}=0$ in
$S(k)$ for long times were marked as cluster-forming (within the coexistence
region). Both in the experiment and simulation, we found that divalent cations
cause phase separation provided the lipid charge is high enough (pH 4.5 or
higher in experiment, $\qpip\le -2$ in simulation).  Monovalent cations were
never seen to induce clusters.

\begin{figure}[t]
\includegraphics[width=86mm]{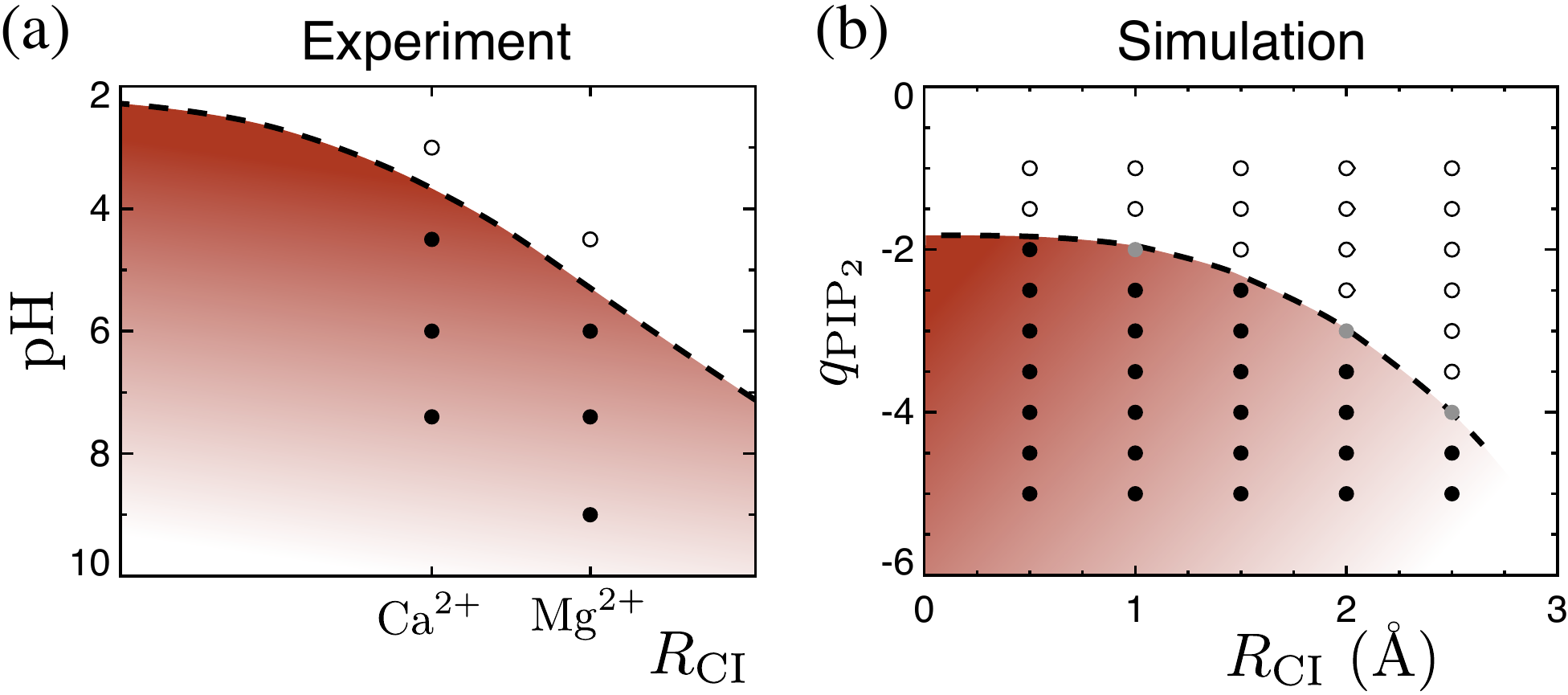}
\caption{Dependence of the minimum magnitude of lipid charge needed for cluster formation on the divalent ion radius.
Colors and symbols are the same as in Fig.~\ref{fig:pd}.
(a) Experimentally ($\pipfrac=0.25$) cluster formation occurs more easily with the smaller of two tested ions (Ca$^{2+}$). (b) The simulation ($\pipfrac=0.05$) shows the same trend with ion radius.}
\label{fig:pdsize}
\end{figure}

Larger divalent ions than \capp\ should mediate weaker attractions,
because larger binding distances imply lower Coulomb
energies.  This effect should manifest itself in a higher charge on
the \pipp\ needed to obtain cluster formation with larger ions. We verified
this in experiments using Mg$^{2+}$, which is known to have a larger hydrated
radius than \capp, even though precise values appear to be
lacking~\footnote{The reported hydrated radii vary from source to source,
mainly due to different methods to determine it, but the literature
consistently reports Mg$^{2+}$ to be larger (3 to 7 \AA) than Ca$^{2+}$ (2.6 to
6.3 \AA)~\cite{ionsizes}.}. We
tentatively indicate this size difference in
Fig.~\ref{fig:pdsize}a.
In agreement with this observation and the theoretical expectation,
the ability of divalent cations to drive cluster formation in our
simulations also decreases with increasing ion size (Fig.~\ref{fig:pdsize}b).
A subtle effect that could influence the effective size
of the ions is the level of dehydration that occurs: the more of the hydration
shell is removed, the closer and hence stronger the binding.   We have therefore
refrained from attempting to assign numerical values to the radii of the
cations in Fig.~\ref{fig:pdsize}a.

\emph{Discussion---}The phase diagram of our model compares surprisingly well
with our experiments without any parameter optimization. The only free
parameter is the dielectric correction factor we apply to compensate for the
underestimation of the mediated binding energy, which we set to $1/3$ from the
start, based on calculations of the electrostatic interactions involved. The
introduction of this correction factor allows a major simplification, of
replacing lipids by spheres of the appropriate charge.  This enabled us to
explore a large parameter space with modest computational resources.
Obviously, this non-systematic coarse-graining approach does not guarantee
quantitative precision, but we note that we obtain quantitative agreement with
experimental values and trends of the surface pressure in the lipid monolayer
before and after adding \capp.  In both simulation and experiment at all
parameter values studied, surface pressures were of order 10 to $30\,$mN/m and
dropped by several mN/m upon addition of \capp\ (data not shown). 

It should be noted that, while hydrogen bonds between the \pipp-molecules exist
and may play a role when the charges are small~\cite{levental08jacs}, our work
strongly suggests that they do not play a dominant role in multivalent
ion-induced clustering --- if they did, having a higher \pipp-charge would make
it harder to form clusters, rather than easier, as we report in
Fig.~\ref{fig:pd}.

Since the interactions in our model have been stripped down to the bare minimum
of electrostatics and steric repulsion,  the only attractive interaction in the
simulations is the Coulomb attraction between the anionic lipids and the
divalent cations.  Therefore, the observed phase separation must be due to
cation-mediated attractions.  The fact that these attractions are the result of
near-cancellation of even larger attractions and repulsions leads to several
special features. We have already mentioned the high energy barriers for lipid
rearrangements that lead to long lived irregular or string-like domain shapes
in our simulations. Such domains are observed experimentally~\cite{levental09}
at pH values close the physiological value (where $\qpip$ is high).  It is
therefore possible that stringlike clusters appear in biological contexts.  

Another striking aspect of ion-mediated attractions is the strong dependence of
the effective attraction strength on the lipid charge. To illustrate this
point, we numerically calculated the binding energy per lipid for a cluster of
30 lipids~\footnote{The reference state for this binding energy is the state in
which they form 15 lipid dimers, neutralized with \capp, so that we can
consider charge neutral clusters and monopole terms will not dominate the
result.}, and found that it increases from $3\,\kT$ at $\qpip=-2$ to $6\,\kT$
at $\qpip=-3$ and $11\,\kT$ at $\qpip=-5$.   At $\qpip=-2$ phase separation
first appears, and at $\qpip=-3$ it is already strong enough to lead to nearly
complete phase separation (see Fig.~\ref{fig:pd}).   This has the potentially
important biological consequence that clusters can form even at the extremely
low concentrations of \pipp\ ($\sim 1\%$) found in the cell membrane.  

We are grateful to Ilya Levental and Alex Travesset for discussions. This work
was supported by the NSF through the Penn MRSEC (DMR-0520021) and NSF-DMR-0605044 (AJL).

\end{document}